\documentclass[twoside,12pt]{article}
\usepackage{CJK}
\usepackage{indentfirst}
\usepackage{bm}
\usepackage{verbatim}
\usepackage{graphicx}
\usepackage{multicol}

\begin{document}
\begin{CJK*}{GBK}{song}
\newcommand{\song}{\CJKfamily{song}}
\newcommand{\hei}{\CJKfamily{hei}}
\newcommand{\fs}{\CJKfamily{fs}}
\newcommand{\kai}{\CJKfamily{kai}}
\def\thefootnote{\fnsymbol{footnote}}

\begin{center}
\Large\hei Discord and entanglement in non-Markovian environments at finite temperature  
\end{center}

\footnotetext{\hspace*{-.45cm}\footnotesize
$^*$Project Supported by the Science and Technology Plan of Hunan Province, China(Grant No. 2010FJ3148) and the National Natural Science Foundation of China(Grant No.11374096).}
\footnotetext{\hspace*{-.45cm}\footnotesize zhmzc1997@126.com, tel:13807314064}

\begin{center}
\rm Hong-Mei Zou, \ Mao-Fa Fang
\end{center}

\begin{center}
\begin{footnotesize} \rm
Key Laboratory of Low-dimensional Quantum Structures and Quantum Control of Ministry of Education, College of Physics and Information Science, Hunan Normal University, Changsha, 410081, China\  \\   

\end{footnotesize}
\end{center}

\vspace*{2mm}

\begin{center}
\begin{minipage}{15.5cm}
\parindent 20pt\footnotesize

The dynamics evolutions of discord and entanglement of two atoms in two independent Lorentzian reservoirs at zero or finite temperature have been investigated by using the time-convolutionless master-equation method. Our results show that, when both the non-Markovian effect and the detuning are present simultaneously, due to the memory and feedback effect of the non-Markovian reservoirs, the discord and the entanglement can be effectively protected even at nonzero temperature by increasing the non-Markovian effect and the detuning. The discord and the entanglement have different robustness for different initial states and their robustness may changes under certain conditions. Nonzero temperature can accelerate the decays of discord and entanglement and induce the entanglement sudden death.

\end{minipage}
\end{center}

\begin{center}
\begin{minipage}{15.5cm}
\begin{minipage}[t]{2.3cm}{\hei Keywords: }\end{minipage}
\begin{minipage}[t]{13.1cm}discord, entanglement, non-Markovian environment, temperature
\end{minipage}\par\vglue8pt
{\bf PACS  }03.65.Yz, 03.67.-a, 42.50.Pq
\end{minipage}
\end{center}

\vspace*{2mm}

\section{\fs Introduction}  
 Entanglement, as a kind of quantum correlation\cite{Schrodinger,Nielsen}, is considered to be not only a vital concept in physics but also a prime resource for quantum information processing\cite{Bouwmeester,Cirac}. Therefore, a great deal of attention has been devoted to the experimental generation and manipulation of entangled systems and the theoretical study of the entanglement evolution\cite{Blinov,Haroche}. In particular, since Yu and Eberly\cite{Eberly} discovered that the Markovian entanglement dynamics of the two qubits exposed to local noisy environments may markedly differ from the single qubit decoherence evolution, it has become an important topic that the analysis of entanglement decay and its relation with decoherence induced by the unavoidable interaction between a system and its environment. Many important progresses have been acquired in experimental and theoretical researches on the entanglement dynamics in open quantum systems. The investigations show that the environmental noise not only can be helpful to protect entanglement but also can make entanglement sudden death(ESD) and entanglement sudden birth(ESB)\cite{mazzola1,zhengh1}.

However, entanglement is not the only type of quantum correlation useful for quantum information processing. Ollivier and Zurek introduced another concept of quantum correlation, termed Discord\cite{Ollivier,Zurek}, which captures the nonclassical correlations, including but not limited to entanglement, that is, separability of the density matrix describing a two-qubit system does not guarantee vanishing of discord. Discord has a significant application in deterministic quantum computation with one pure qubit(DQC1)\cite{Datta} and estimation of quantum correlations in the Grover search algorithm\cite{CuiJ}. On the other hand, discord has also been extensively used in studies of quantum phase transition \cite{Dillenschneider,Sarandy} and to measure the quantum correlation between relatively accelerated observers\cite{Datta2}.

In fact, any realistic physical system will suffer from unwanted interactions with the outside environments
\cite{Breuer}, causing decoherence and destroying entanglement and discord. In recent years, the discord and the entanglement for open quantum systems have been widely studied by means of different methods. For instances, Bellomo B \emph{et al} studied the entanglement dynamics of two independent qubits in non-Markovian environments\cite{Bellomo1,Bellomo2}. Xiao X \emph{et al}\cite{XiaoX} discussed the discord dynamics of two qubits in this model. In Refs\cite{Bellomo2,XiaoX,WangB,Werlang}, the same model is adopted, in which the whole system can be divided into two parts and a formal solution for the evolution may be obtained by exactly solving the time-dependent Schr\"{o}dinger equation in the subspace with one excitation.
The results show that discord is much more robust compared to entanglement in open quantum systems. In this paper, we will apply another method to handle this model, where all the parts are included as a whole system and a numerical simulation can be obtained by using the time-convolutionless(TCL) master-equation method, and investigate the influences of the non-Markovian effect, the detuning and the temperature as well as atomic initial states on the discord and the entanglement.

The report is structured as follows. We present a physical model in section 2. In the section 3, we introduce discord and entanglement of a two-qubit system. Results and discussions are given in section 4. We conclude briefly the report in section 5.

\section{\fs Physical Model}
We consider two two-level atoms, where each atom couples with a structured reservoir\cite{Ferraro}. The Hamiltonian is($\hbar=1$)
$H=H_{0}+\alpha H_{I}$, where $H_{0}=\omega_{0}\sum_{j=A}^{B}S_{j}^{z}+\sum_{n}\omega_{n,A}b_{n,A}^{\dag}b_{n,A}+\sum_{m}\omega_{m,B}b_{m,B}^{\dag}b_{m,B}$
is the free Hamiltonian of the combined system. $\omega_{0}$ is the transition frequency of the atom, $S_{j}^{z}$ is the inversion operators describing the atom $j$($j=A$ or $B$), $b_{n,A}^{\dag}(b_{m,B}^{\dag})$ and $b_{n,A}(b_{m,B})$ are the creation and annihilation operators of the reservoir with the frequency $\omega_{n,A}(\omega_{m,B})$. The parameter $\alpha$ is a dimensionless expansion parameter. In the interaction picture, the Hamiltonian $\alpha H_{I}$ reads $\alpha H_{I}(t)=\sum_{j=A}^{B}(S_{j}^{+}\sum_{n}g_{n,j}b_{n,j}e^{i(\omega_{0}-\omega_{n,j})t}+h.c.)$,
where $g_{n,j}$ is the coupling constant between the atom and its reservoir, $S_{j}^{+}$ and $S_{j}^{-}$ are the upward and downward operators of the atom, respectively.

In the second order approximation, the TCL master equation\cite{Breuer} of $\rho_{AB}(t)$ has the form
\begin{equation}\label{EB04}
\frac{d}{dt}\rho_{AB}{(t)}=-\alpha^{2}\int_{0}^{t}d\tau Tr_{E}([H_{I}{(t)},[H_{I}{(\tau)},\rho_{AB}{(t)}\otimes\rho_{E}]])
\end{equation}
with the environment state $\rho_{E}$. Here we have supposed that $\rho{(t)}=\rho_{AB}{(t)}\otimes\rho_{E}$ and $Tr_{E}([H_{I}{(t)}$,\\
$\rho_{AB}{(0)}\otimes\rho_{E}])=0$, thus the TCL master equation may be written as
\begin{eqnarray}\label{EB06}
\frac{d}{dt}\rho_{AB}{(t)}&=&\mathcal{L}^{(A)}\rho_{AB}{(t)}+\mathcal{L}^{(B)}\rho_{AB}{(t)},
\end{eqnarray}
where $\mathcal{L}^{(j)}$ is the Liouville super-operator\cite{ZouHM} and it is defined by
\begin{eqnarray}\label{EB07}
\mathcal{L}^{(j)}\rho_{AB}{(t)}&=&f_{j}(t)[S_{j}^{-}\rho_{AB}{(t)},S_{j}^{+}]+f_{j}^\ast(t)[S_{j}^{-},\rho_{AB}{(t)}S_{j}^{+}]\nonumber\\
&&+k_{j}^\ast(t)[S_{j}^{+}\rho_{AB}{(t)},S_{j}^{-}]+k_{j}(t)[S_{j}^{+},\rho_{AB}{(t)}S_{j}^{-}].
\end{eqnarray}
The correlation functions $k_{j}(t)$ and $f_{j}(t)$ are given by
\begin{eqnarray}\label{EB08}
k_{j}(t)&=&i\sum_{n}|g_{n,j}|^{2}\langle b_{n,j}^{\dagger}b_{n,j}\rangle_{E_{j}}\frac{1-e^{i(\omega_{0}-\omega_{n,j})t}}{\omega_{0}-\omega_{n,j}}
\end{eqnarray}
and
\begin{eqnarray}\label{EB09}
f_{j}(t)&=&i\sum_{n}|g_{n,j}|^{2}\langle b_{n,j}b_{n,j}^{\dagger}\rangle_{E_{j}}\frac{1-e^{i(\omega_{0}-\omega_{n,j})t}}{\omega_{0}-\omega_{n,j}},
\end{eqnarray}
where $\langle O\rangle_{E_{j}}\equiv Tr_{E_{j}}(O\rho_{E_{j}})$.

Assuming that the two reservoirs are identical and initially prepared in a thermal state with temperature $T$, the correlation functions reduce to
\begin{eqnarray}\label{EB10}
k(t)&=&i\sum_{n}|g_{n}|^{2}\frac{1}{e^{\hbar\omega_{n}/k_{B}T}-1}\frac{1-e^{i(\omega_{0}-\omega_{n})t}}{\omega_{0}-\omega_{n}},\nonumber\\
f(t)&=&i\sum_{n}|g_{n}|^{2}\frac{e^{\hbar\omega_{n}/k_{B}T}}{e^{\hbar\omega_{n}/k_{B}T}-1}\frac{1-e^{i(\omega_{0}-\omega_{n})t}}{\omega_{0}-\omega_{n}},
\end{eqnarray}
here $k_{B}$ is the Boltzmann constant. For a sufficiently large environment, we can replace the sum over the discrete coupling constants with an integral over a continuous distribution of frequencies of the environmental modes, \emph{i.e.} $\sum_{n}|g_{n}|^{2}\rightarrow\int_{0}^{\infty}d\omega J(\omega)$.

Let the spectral density has a Lorentzian form
\begin{equation}\label{EB12}
J(\omega)=\frac{1}{2\pi}\frac{\gamma_{0}\lambda^{2}}{(\omega_{0}-\omega-\delta)^{2}+\lambda^{2}},
\end{equation}
where $\delta$ is the detuning between $\omega_{0}$ and the reservoir center frequency $\omega$. The parameter $\lambda$ defines the spectral width of the coupling, which is connected to the reservoir correlation time $\tau_{R}$ by $\tau_{R}$=$\lambda^{-1}$ and the parameter $\gamma_{0}$ is related to the relaxation time scale $\tau_{S}$ by $\tau_{S}$=$\gamma_{0}^{-1}$. In the subsequent analysis, typically a weak and a strong coupling regimes can be distinguished. For a weak regime, $\lambda>2\gamma_{0}$(i.e. $\tau_{S}>2\tau_{R}$), the dynamical behavior of the system is essentially a Markovian exponential decay controlled by $\gamma_{0}$. In the strong coupling regime, $\lambda<2\gamma_{0}$(i.e. $\tau_{S}<2\tau_{R}$), the non-Markovian effects become relevant\cite{ZouHM2}.

 Inserting Eq.~(\ref{EB12}) to Eqs.~(\ref{EB10}), we obtain the correlation functions
\begin{eqnarray}\label{EB13}
k(t)&=&\frac{\gamma_{0}\lambda(1-e^{(i\delta-\lambda)t})}{2(\lambda-i\delta)}\frac{1}{e^{\hbar(\omega_{0}-\delta-i\lambda)/k_{B}T}-1},\nonumber\\
f(t)&=&\frac{\gamma_{0}\lambda(1-e^{(i\delta-\lambda)t})}{2(\lambda-i\delta)}\frac{e^{\hbar(\omega_{0}-\delta-i\lambda)/k_{B}T}}{e^{\hbar(\omega_{0}-\delta-i\lambda)/k_{B}T}-1}
\end{eqnarray}

\section{\fs Discord and Entanglement}
For any two-qubit system $\rho_{AB}$, its total correlation can be written as $\mathcal{I}(\rho_{AB})=S(\rho_{A})+S(\rho_{B})-S(\rho_{AB})$, where $S(\rho)=-Tr(\rho log_{2}\rho)$ is the von Neumann entropy\cite{Nielsen} of $\rho$. Its classical correlation is $\mathcal{C}(\rho_{AB})=max_{\{\Pi_{k}\}}[S(\rho_{A})$
$-S(\rho_{A|B})]$, here the maximum represents the most information gained about subsystem $A$ as a result of the perfect measurement $\{\Pi_{k}\}$ on subsystem $B$. $S(\rho_{A|B})=\sum_{k=1}^{2}p^{k}S(\rho^{k}_{A})$ is the quantum conditional entropy of subsystem $A$: $p^{k}_{A}=Tr_{AB}(\Pi_{k}\rho_{AB}\Pi_{k})$ and $\rho^{k}_{A}=Tr_{B}(\Pi_{k}\rho_{AB}\Pi_{k})/p^{k}_{A}$ are the probability and the state of subsystem $A$ for measurement outcome $k$. Discord is defined as the difference between the total correlation ($\mathcal{I}(\rho_{AB})$) and the classical correlation ($\mathcal{C}(\rho_{AB})$), i.e.
\begin{eqnarray}\label{EB17}
\mathcal{D}(\rho_{AB})&=&\mathcal{I}(\rho_{AB})-\mathcal{C}(\rho_{AB})
\end{eqnarray}
and is interpreted as a measurement of quantum correlations\cite{Ollivier}. For X state, described by the following density matrix,
\begin{eqnarray}\label{EB18}
\rho_{AB}&=&\left(
            \begin{array}{cccc}
              \rho_{11} & 0 & 0 & \rho_{14} \\
              0& \rho_{22}& \rho_{23}&0 \\
              0& \rho_{32}& \rho_{33} &0 \\
              \rho_{41} &0&0& \rho_{44}\\
            \end{array}
          \right),
\end{eqnarray}
Discord can be calculated\cite{WangCZ}, analytically, as
\begin{eqnarray}\label{EB19}
\textit{D}(\rho_{AB})&=&min\{{Q}_{1},{Q}_{2}\},
\end{eqnarray}
where ${Q}_{j}=h(\rho_{11}+\rho_{33})+\sum_{k=1}^{4}\lambda_{k}log_{2}\lambda_{k}+D_{j}$, with $\lambda_{k}$ being the eigenvalues of $\rho_{AB}$ and $h(x)=-xlog_{2}x-(1-x)log_{2}(1-x)$ is the binary entropy. Here $D_{1}=h(\tau)$, where $\tau=\frac{1}{2}\{1+\sqrt{[1-2(\rho_{33}+\rho_{44})]^{2}+4(|\rho_{14}|+|\rho_{23}|)^{2}}\}$ and $D_{2}=-\sum_{k=1}^{4}\rho_{kk}log_{2}\rho_{kk}-h(\rho_{11}+\rho_{33})$.

In addition to discord, another quantum correlation is entanglement, which can be calculated through entanglement of formation (EoF), as\cite{Wootter}
\begin{equation}\label{EB20}
\textit{E}(\rho_{AB})=H(\frac{1+\sqrt{1-C^{2}}}{2})
\end{equation}
where $H(x)=-xlog_{2}x-(1-x)log_{2}(1-x)$, $C=max(0,\sqrt{\lambda_{1}}-\sqrt{\lambda_{2}}-\sqrt{\lambda_{3}}-\sqrt{\lambda_{4}})$ is Wootter's concurrence and $\lambda_{i}$ are the eigenvalues, organized in a descending order, of the matrix $\tilde{\rho}
=\rho(\sigma_{y}\otimes\sigma_{y})\rho^{\ast}(\sigma_{y}\otimes\sigma_{y})$. For the X state such as Eq.~(\ref{EB18}), the concurrence of $\rho_{AB}$ is $C=2max(0,|\rho_{14}|-\sqrt{\rho_{22}\rho_{33}},|\rho_{23}|-\sqrt{\rho_{11}\rho_{44}})$\cite{mazzola1}.

Discord is equal to entanglement for pure states, but their relation is complicated for the mixed states, that is, there are separable mixed states with nonzero discord\cite{Ferraro2}. In the following, our main concern will be the dynamics behaviors of discord and entanglement of two atoms in Lorentzian reservoirs at zero and finite temperature.

\section{\fs Results and Discussions}
In the section, we discuss in detail the dynamic behaviors of discord and entanglement of two atoms in Lorentzian reservoirs. If the initial atomic states are
\begin{eqnarray}\label{EB22}
|\psi(0)\rangle=\frac{1}{2}(|00\rangle+|11\rangle), |\phi(0)\rangle=\frac{1}{2}(|01\rangle+|10\rangle),
\end{eqnarray}
the atomic density matrix $\rho_{AB}(t)$ can remain X structure under the master equation Eq.~(\ref{EB06}). Utilizing Eq.~(\ref{EB19}) and Eq.~(\ref{EB20}), we can calculate the discord($D$) and the entanglement($E$) of $\rho_{AB}(t)$. In the following discussions, we first give the dynamics evolutions of discord and entanglement in the resonant case, then we analyze those in the non-resonant case.

\subsection{\fs Resonant case}
We assume that the two atoms couple resonantly to the Lorentzian reservoirs(i.e. $\delta=0$) and study the influence of the temperature, the non-Markovian effect and the initial states on the discord and the entanglement.

\emph{Case I}. The Markovian regime

Fig.1 presents the dynamics of discord and entanglement for different initial states and different temperatures when $\lambda=5\gamma_{0}$ and $\delta=0$.

Fig.1(a) displays the dynamics of discord and entanglement at $T=0$ for $|\psi(0)\rangle$. The results show that, both the discord and the entanglement decay exponentially and vanish asymptotically, and their robustness can change at the critical time, which is expressed by $t_{cri}$ and $\gamma_{0}t_{cri}=0.32$, namely, the discord reduces faster slightly than the entanglement when $\gamma_{0}t<0.32$ while the discord is more robust than the entanglement when $\gamma_{0}t>0.32$. Fig.1(b) gives the dynamics of discord and entanglement at $T=0$ for $|\phi(0)\rangle$. Comparing Fig.1(a) and Fig.1(b), we find that the discord for $|\phi(0)\rangle$ is a little robust than that for $|\psi(0)\rangle$, and likewise for the entanglement, and $\gamma_{0}t_{cri}=0.6>0.32$, namely, the $t_{cri}$ for $|\phi(0)\rangle$ is bigger than that for $|\psi(0)\rangle$.

When $\frac{k_{B}T}{\hbar\omega_{0}}=1$, the dynamics behaviors of discord and entanglement for $|\psi(0)\rangle$ and $|\phi(0)\rangle$ are respectively plotted in Fig.1(c) and 1(d). In Fig.1(c), there is $\gamma_{0}t_{cri}=0.22$, and it is worth noting that the discord still reduces quickly and asymptotically to zero but the entanglement can vanish completely in a very short time, i.e. ESD, whose time is expressed by $t_{ESD}$ and $\gamma_{0}t_{ESD}=0.73$. In Fig.1(d), there are $\gamma_{0}t_{cri}=0.22$ and $\gamma_{0}t_{ESD}=0.78$, so the entanglement behavior in Fig.1(d) is very similar to that in Fig.1(c) except that the entanglement decay in Fig.1(d) is a little slower than that in Fig.1(c), and likewise for the discord. Namely, the initial states hardly affect the discord and the entanglement at nonzero temperature, which is markedly different from the zero temperature case.

\begin{center}
\includegraphics[width=14cm,height=10cm]{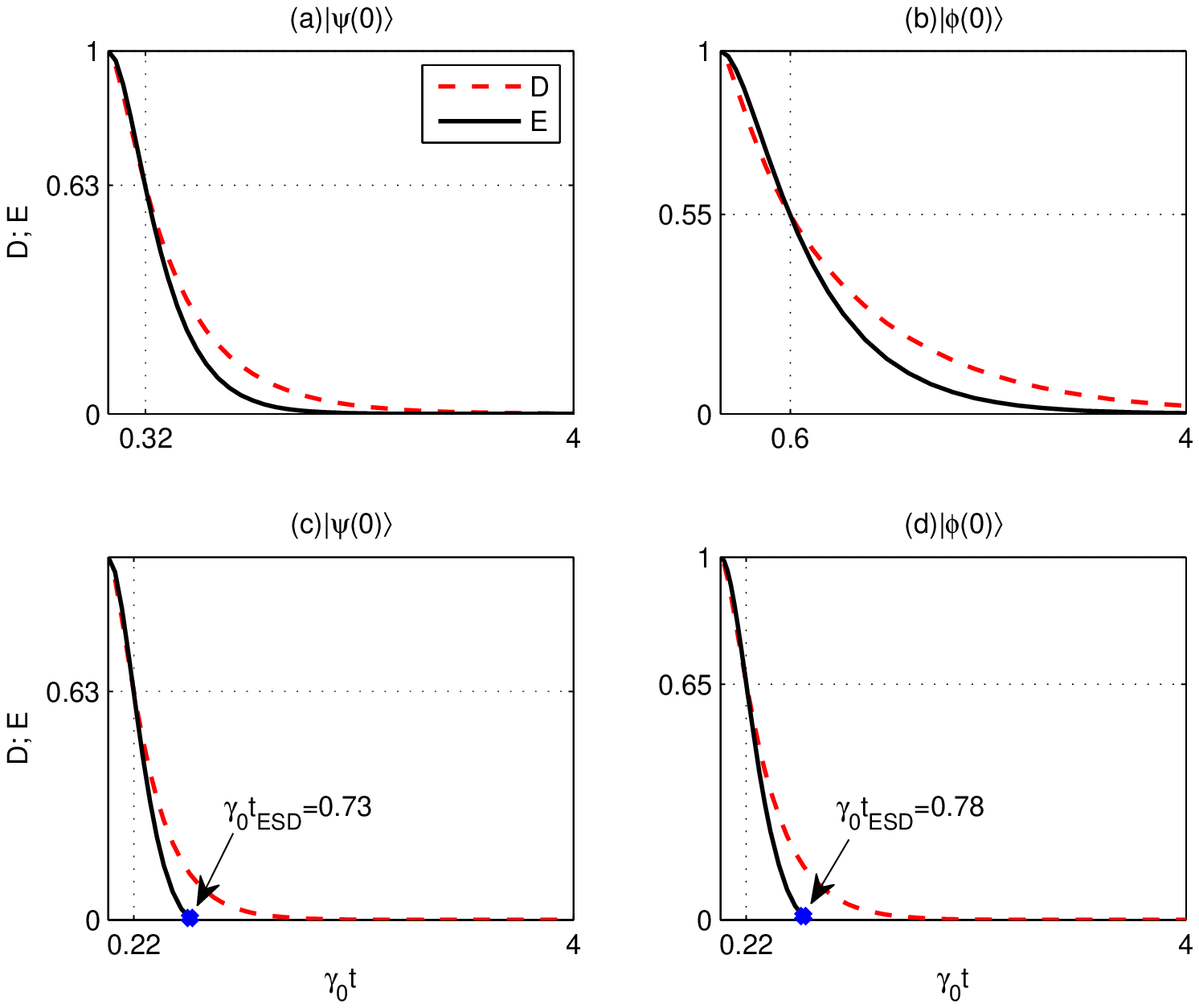}
\parbox{14cm}{\small{\bf Fig1.}
For different initial states and different temperatures, $D$ and $E$ versus $\gamma_{0}t$ when $\delta=0$ and $\lambda=5\gamma_{0}$. $D$(dashed line), $E$(solid line). (a),(b)$T=0$. (c),(d)$\frac{k_{B}T}{\hbar\omega_{0}}=1$. Other $\omega_{0}=10\gamma_{0}$.}
\end{center}

Comparing Fig.1(c)-(d) and Fig.1(a)-(b) respectively, we can see that, nonzero temperature can accelerate the decays of the discord and the entanglement and shorten the $t_{cri}$. More important, nonzero temperature can yet induce the ESD. The reason is that the thermal effect of environment can enhance the atomic dissipation so that the discord and the entanglement reduce more quickly and the discord is more robust against temperature than the entanglement.

Therefore, under the resonance and the Markovian regime, the discord for $|\phi(0)\rangle$ is more robust than that for $|\psi(0)\rangle$, and likewise for the entanglement. Nonzero temperature can accelerate the decays of the discord and the entanglement and induce the ESD.

\emph{Case II}. The non-Markovian regime

Fig.2 depicts the evolutions of discord and entanglement versus $\gamma_{0}t$ for different initial states and different temperatures when $\lambda=0.1\gamma_{0}$ and $\delta=0$.

Fig.2(a) shows that, for $|\psi(0)\rangle$, both the discord and the entanglement at $T=0$ decay exponentially and vanish asymptotically after a short time, and $\gamma_{0}t_{cri}=1.84$. Fig.2(b) describes that, for $|\phi(0)\rangle$, the discord and the entanglement at $T=0$ will also reduce asymptotically to zero and $\gamma_{0}t_{cri}=3.0$. Comparing Fig.2(b) and Fig.2(a), we find that, the discord for $|\phi(0)\rangle$ is robust than that for $|\psi(0)\rangle$, and likewise for the entanglement. Fig.2(c) and Fig.2(d) respectively give the discord and the entanglement for $|\psi(0)\rangle$ and $|\phi(0)\rangle$ at $\frac{k_{B}T}{\hbar\omega_{0}}=1$. Fig.2(c) tells us that, $\gamma_{0}t_{cri}=1.2$ and $\gamma_{0}t_{ESD}=3.0$. In Fig.2(d), there are $\gamma_{0}t_{cri}=1.2$ and $\gamma_{0}t_{ESD}=3.1$, so the entanglement behavior in Fig.2(d) is very similar to that in Fig.2(c) except that the decay of the entanglement in Fig.2(d) is a little slower than that in Fig.2(c), and likewise for the discord. Comparing Fig.2(c)-(d) and Fig.2(a)-(b) respectively, it can be observed that, nonzero temperature can also accelerate the decays of the discord and the entanglement and induce the ESD.

\begin{center}
\includegraphics[width=14cm,height=10cm]{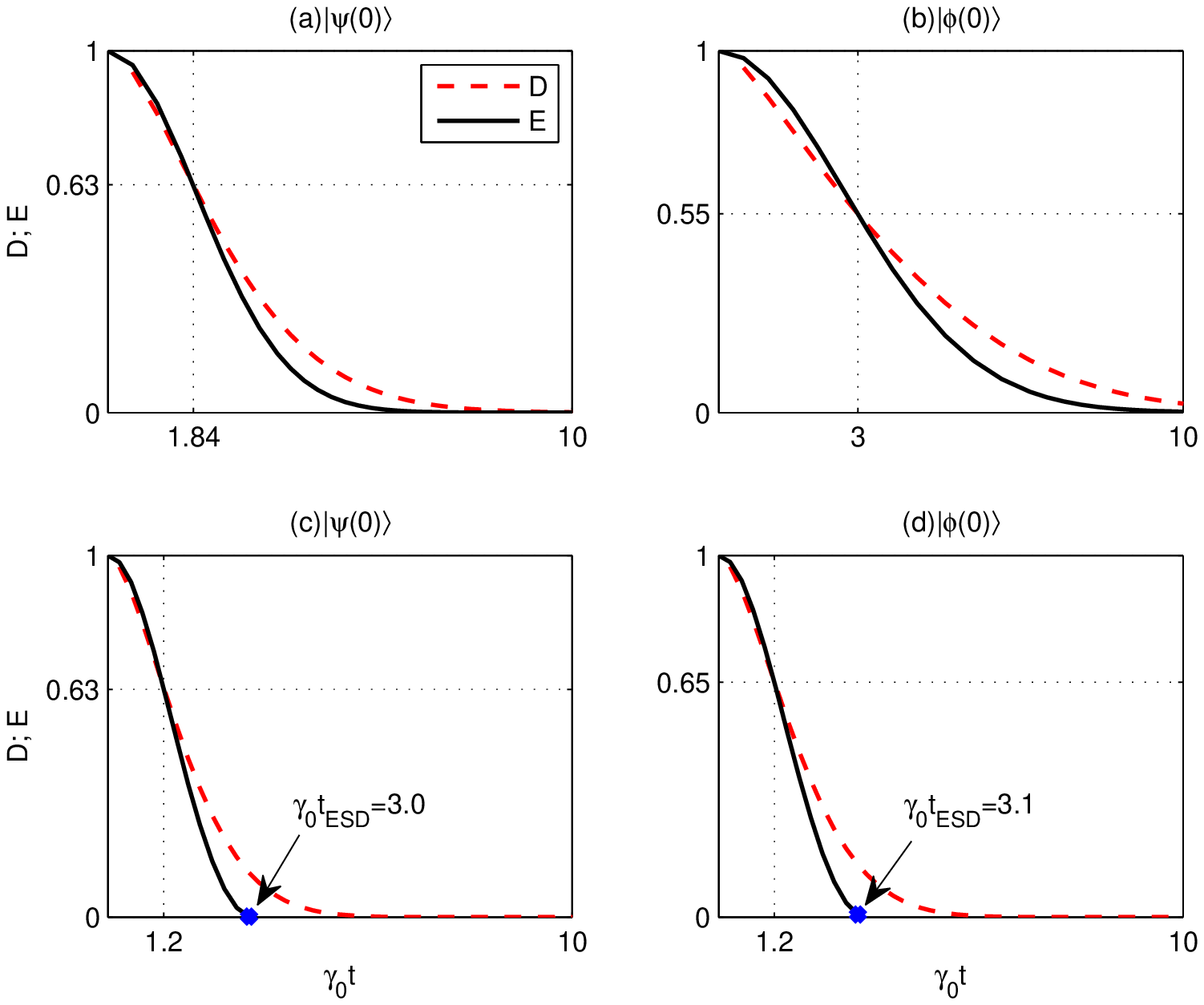}
\parbox{14cm}{\small{\bf Fig2.}
For different initial states and different temperatures, $D$ and $E$ versus $\gamma_{0}t$ when $\lambda=0.1\gamma_{0}$ and $\delta=0$. $D$(dashed line), $E$(solid line). (a),(b)$T=0$. (c),(d)$\frac{k_{B}T}{\hbar\omega_{0}}=1$. Other $\omega_{0}=10\gamma_{0}$.}
\end{center}

Comparing Fig.2(a)-(d) and Fig.1(a)-(d) respectively, we know that, in the resonance, the non-Markovian effect can reduce the decay rates of the discord and the entanglement and prolong $t_{cri}$ and $t_{ESD}$, but these influences are very small.

\subsection{\fs Non-resonant case}
Above, we have restricted our discussions to the resonant case, but the detuning case (i.e. $\delta\neq\gamma_{0}$) would be more reasonable.

\emph{Case I}. The Markovian regime

Fig.3 displays the dynamics of discord and entanglement for different initial states and different temperatures when $\lambda=5\gamma_{0}$ and $\delta=\gamma_{0}$. Comparing Fig.3 and Fig.1, we see that, the decay rates of discord in Fig.3(a)-(d) are approximately equal to those in Fig.1(a)-(d) respectively, and likewise for the entanglement. But, the $t_{ESD}$ in Fig.3(c)-(d) is a little bigger than that in Fig.1(c)-(d), respectively. As a result, in the Markovian regime, the detuning hardly impacts on the discord and the entanglement.

\begin{center}
\includegraphics[width=14cm,height=10cm]{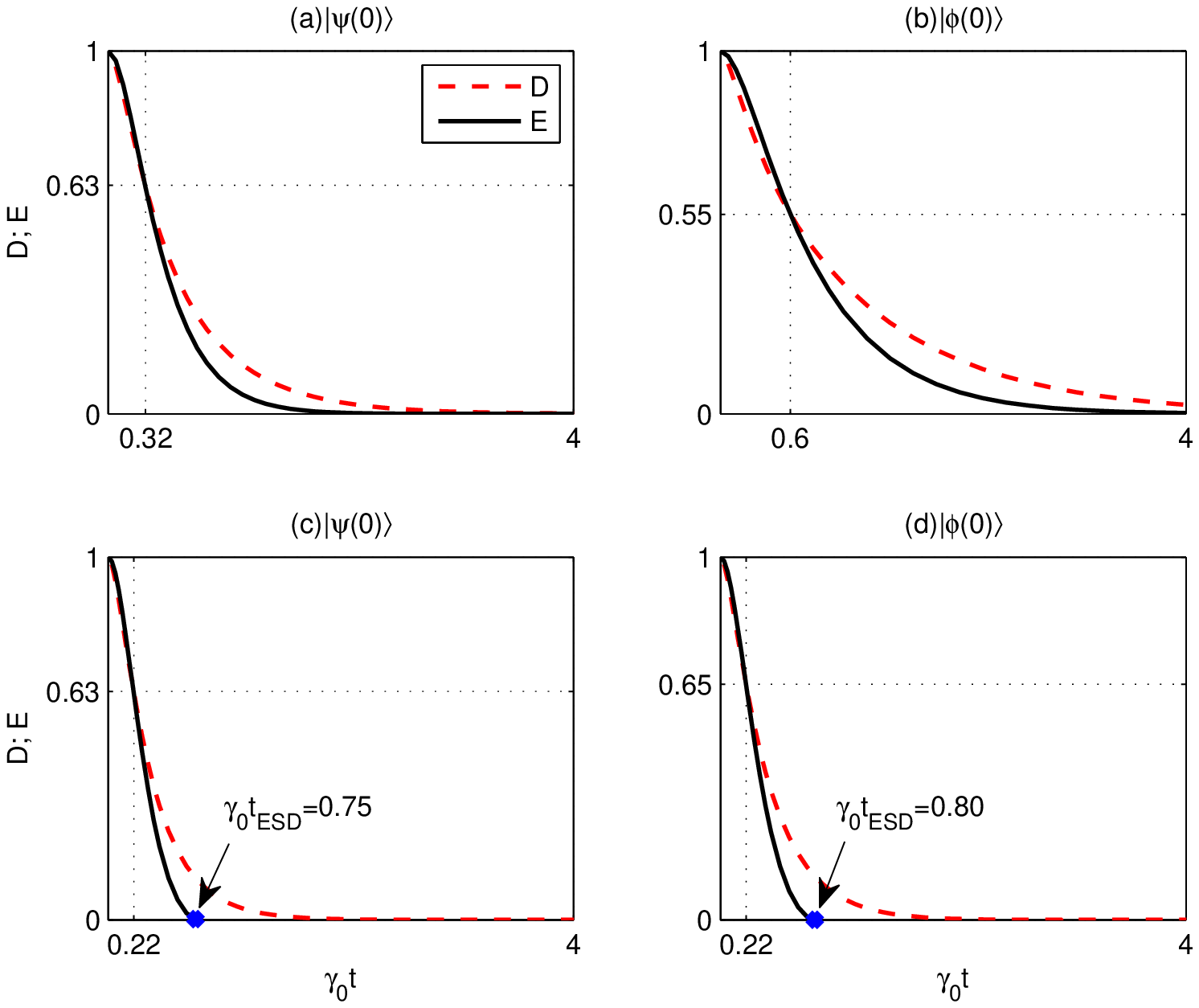}
\parbox{14cm}{\small{\bf Fig3.}
For different initial states and different temperatures, $D$ and $E$ versus $\gamma_{0}t$ when $\lambda=5\gamma_{0}$ and $\delta=\gamma_{0}$. $D$(dashed line), $E$(solid line). (a),(b)$T=0$. (c),(d)$\frac{k_{B}T}{\hbar\omega_{0}}=1$. Other $\omega_{0}=10\gamma_{0}$.}
\end{center}

\emph{Case II}. The non-Markovian regime

Fig.4 exhibits the influence of initial states and temperatures on the discord and the entanglement when both the non-Markovian effect and the detuning are present simultaneously(i.e. $\lambda=0.1\gamma_{0}$ and $\delta=\gamma_{0}$).

Fig.4(a) gives the dynamics behaviors of discord and entanglement for $|\psi(0)\rangle$ at $T=0$. In this case, due to the memory and feedback effect of reservoirs, both the discord and the entanglement are obviously protected, but they will oscillate damply then reduce asymptotically to zero in a long time(see the inset in Fig.4(a)) and $\gamma_{0}t_{cri}=7.7$. The relations of the discord and the entanglement versus $\gamma_{0}t$ for $|\phi(0)\rangle$ at $T=0$ are plotted in Fig.4(b). It is seen that, $\gamma_{0}t_{cri}=31$, and both the discord and the entanglement are also obviously protected. Comparing Fig.4(b) and Fig.4(a), it is very evidently that the discord for $|\phi(0)\rangle$ reduces much slower than that for $|\psi(0)\rangle$, and likewise for the entanglement.

In Fig.4(c) and Fig.4(d), we draw the dynamics of discord and entanglement for $|\psi(0)\rangle$ and $|\phi(0)\rangle$ at $\frac{k_{B}T}{\hbar\omega_{0}}=1$. Fig.4(c) shows that, for $|\psi(0)\rangle$, there is $\gamma_{0}t_{cri}=1.2$ and the discord and the entanglement all oscillate damply then the discord decrease asymptotically to zero but the entanglement will be sudden death at $\gamma_{0}t_{ESD}=33$. In Fig.4(d), there are $\gamma_{0}t_{cri}=1.2$ and $\gamma_{0}t_{ESD}=34$, that is, the discord behavior for $|\phi(0)\rangle$ is very similar to that for $|\psi(0)\rangle$, and likewise for the entanglement. Comparing Fig.4(c)-(d) and Fig.4(a)-(b) respectively, it can shown that, nonzero temperature can also accelerate the decays of discord and entanglement, enlarge the oscillating amplitudes of discord and entanglement and induce the ESD even if both the non-Markovian effect and the detuning are present simultaneously.

\begin{center}
\includegraphics[width=14cm,height=10cm]{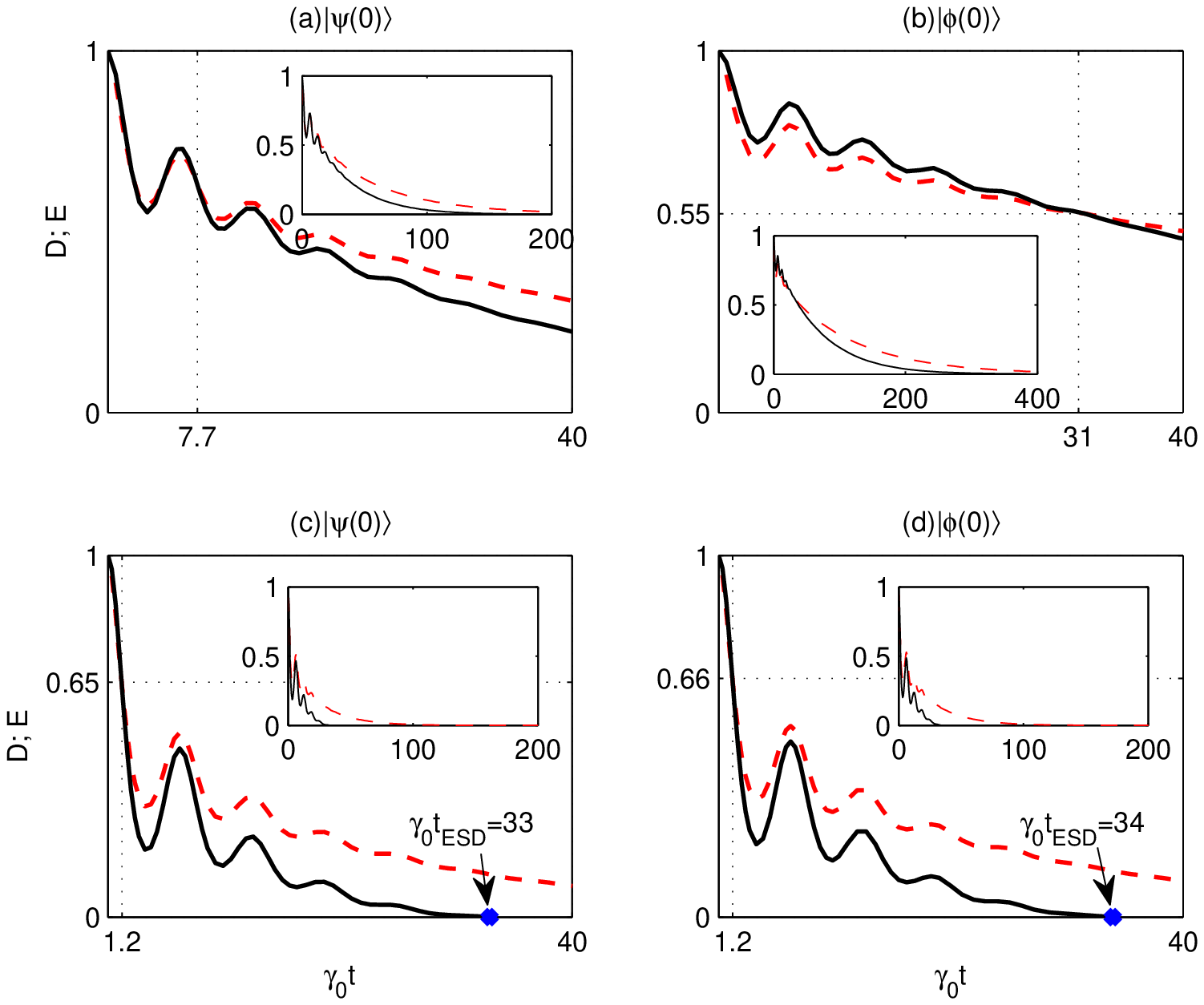}
\parbox{14cm}{\small{\bf Fig4.}
The influence of initial states and temperatures on $D$ and $E$ when $\lambda=0.1\gamma_{0}$ and $\delta=\gamma_{0}$. $D$(dashed line), $E$(solid line). (a),(b)$T=0$. (c),(d)$\frac{k_{B}T}{\hbar\omega_{0}}=1$. The insets show the long-time behaviors. Other $\omega_{0}=10\gamma_{0}$.}
\end{center}

Comparing Fig.4 with Fig.2 and Fig.3, we find that there is essentially different. If there is only the non-Markovian effect or only the detuning, the discord and the entanglement will monotonously and rapidly decline to zero. However, if the non-Markovian effect and the detuning are present simultaneously, the discord and the entanglement will oscillate damply and their decay rates also become evidently smaller. The physical interpretation is that, when only $\lambda<2\gamma_{0}$(the non-Markovian effect) or only $\delta\neq0$(the detuning), the quantum information can flow out from the atomic system but is hardly returned from the reservoirs so that the discord and the entanglement monotonously and rapidly reduce, as shown in Fig.2 and Fig.3. However, when $\lambda<2\gamma_{0}$ and $\delta\neq0$, due to the memory and feedback effect of the non-Markovian reservoir, the quantum information flowing to the reservoir will be partly returned to the atom so that the discord and the entanglement oscillate damply and their decay rates become smaller, as shown in Fig.4.

Fig.5 reveals the dynamics of discord and entanglement when $\lambda=0.1\gamma_{0}$ and $\delta=4\gamma_{0}$. From Fig.5(a), we see that, for $|\psi(0)\rangle$ and $T=0$, the discord and the entanglement can be effectively protected and the entanglement is more robust than the discord although they can reduce asymptotically to zero in a very long time(see the inset in Fig.5(a)). Fig.5(b) shows that the discord for $|\phi(0)\rangle$ is more robust than that for $|\psi(0)\rangle$, and likewise for the entanglement. Fig.5(c) states clearly that, the discord and the entanglement can still be effectively protected at $\frac{k_{B}T}{\hbar\omega_{0}}=1$, but their decay rates and oscillating amplitudes are a little bigger than those at $T=0$. The entanglement in Fig.5(d) is similar to that in Fig.5(c) expect that the decay rate in Fig.5(d) is a little smaller than that in Fig.5(c), and likewise for the discord.

\begin{center}
\includegraphics[width=14cm,height=10cm]{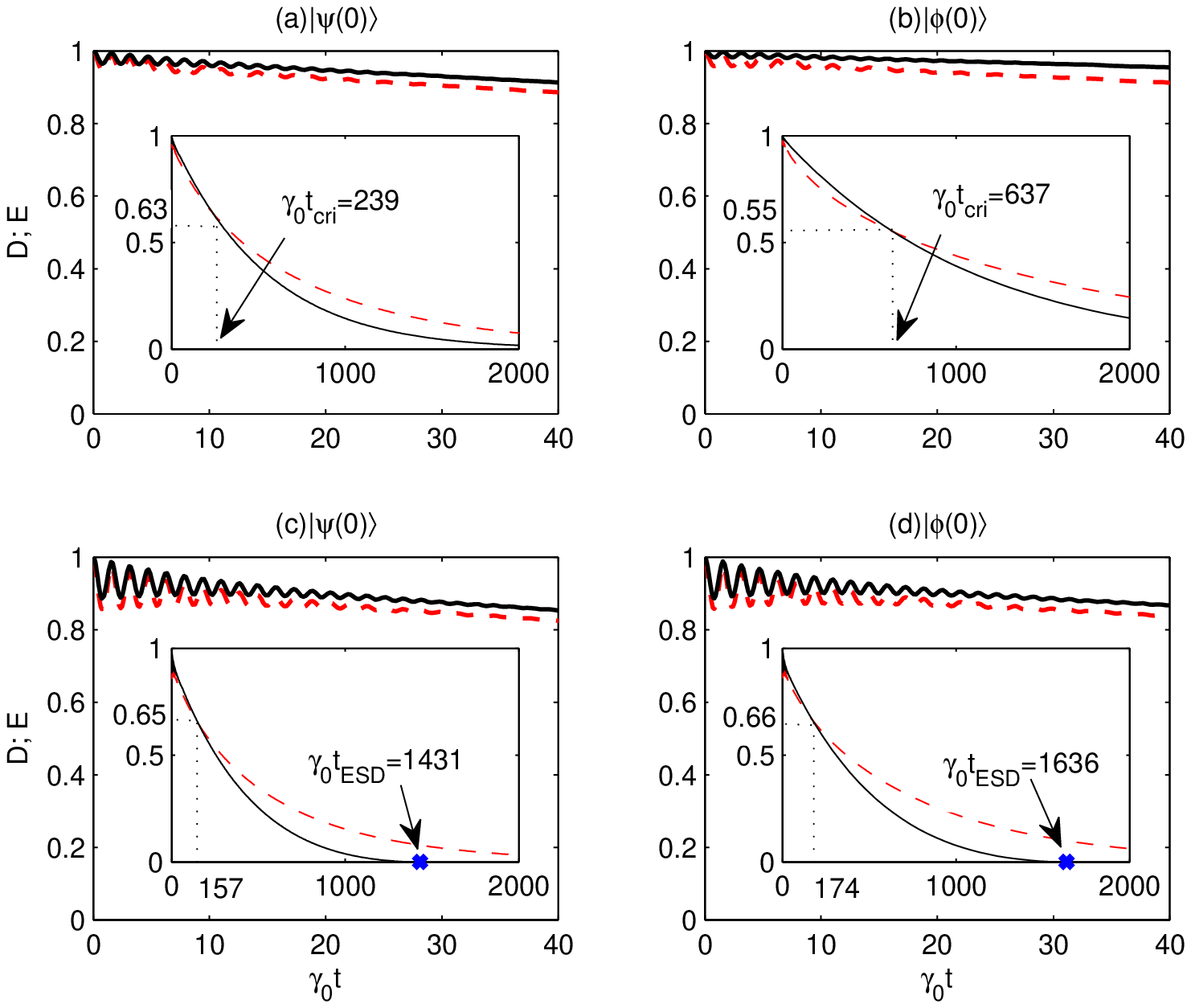}
\parbox{14cm}{\small{\bf Fig5.}
The dynamics behaves of $D$ and $E$ in the non-Markovian regime($\lambda=0.1\gamma_{0}$) and with the big detuning($\delta=4\gamma_{0}$). $D$(dashed line), $E$(solid line). (a),(b)$T=0$. (c),(d)$\frac{k_{B}T}{\hbar\omega_{0}}=1$. The insets show the long-time dynamics. Other $\omega_{0}=10\gamma_{0}$.}
\end{center}

Comparing Fig.5(a)-(d) and Fig.4(a)-(d) respectively, we know that, in the non-Markovian regime, increasing the detuning can prolong $t_{cri}$ and $t_{ESD}$(see the insets in Fig.5), enlarge the oscillating frequency and reduce the decays of discord and entanglement. More important, even at nonzero temperature, the discord and the entanglement can also be effectively protected by means of the big detuning. The physical explanation is that, in the non-Markovian regime, with $\delta$ increasing, the atom can more quickly exchange the quantum information with its reservoir so that the atomic decay becomes slower and the quantum information returned from the reservoir is more, hence the discord and the entanglement can be more effectively protected.

Fig.6 exhibits the dynamics of discord and entanglement when $\lambda=0.01\gamma_{0}$ and $\delta=\gamma_{0}$. From Fig.6(a), we see that, for $|\psi(0)\rangle$ and $T=0$, the discord and the entanglement can be also effectively protected although the discord and the entanglement yet reduce asymptotically to zero in a very long time(see the inset in Fig.6(a)). Fig.6(b) reveals that, for $|\phi(0)\rangle$ and $T=0$, the discord for $|\phi(0)\rangle$ is more robust than that for $|\psi(0)\rangle$, and likewise for the entanglement. Fig.6(c) shows that, the discord and the entanglement can also be effectively protected at $\frac{k_{B}T}{\hbar\omega_{0}}=1$, but their decay rates and oscillating amplitudes are a little bigger than those at $T=0$. The entanglement in Fig.6(d) is very similar to that in Fig.6(c) expect that the decay rate in Fig.6(d) is a little smaller than that in Fig.6(c), and likewise for the discord.

Comparing Fig.6(a)-(d) and Fig.4(a)-(d) respectively, we know that, under the detuning, increasing the non-Markovian effect can prolong $t_{cri}$ and $t_{ESD}$(see the insets in Fig.6) and reduce the decays of discord and entanglement. More important, even at nonzero temperature, the discord and the entanglement can also be effectively protected by means of the strong non-Markovian effect. The physical explanation is that, under the detuning, with $\lambda$ decreasing, the memory and feecback effect of the non-Markovian reservoir can become strong so that the atomic decay becomes slower and the quantum information returned from the reservoir is more, hence the discord and the entanglement can be more effectively protected.

\begin{center}
\includegraphics[width=14cm,height=10cm]{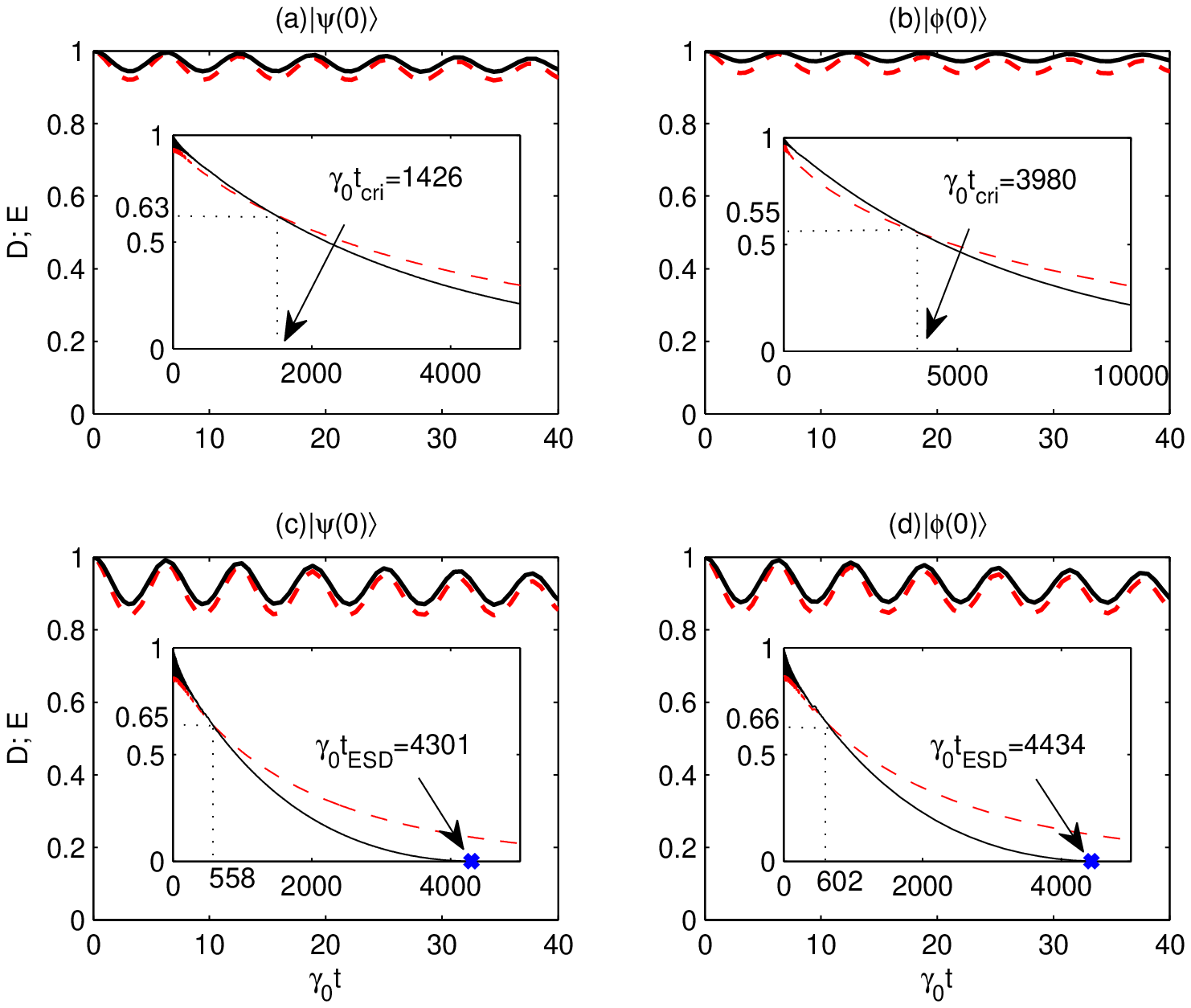}
\parbox{14cm}{\small{\bf Fig6.}
The dynamics behaves of $D$ and $E$ with the strong non-Markovian effect($\lambda=0.01\gamma_{0}$) and the small detuning($\delta=\gamma_{0}$). $D$(dashed line), $E$(solid line). (a),(b)$T=0$. (c),(d)$\frac{k_{B}T}{\hbar\omega_{0}}=1$. The insets show the long-time dynamics. Other $\omega_{0}=10\gamma_{0}$.}
\end{center}

\section{\fs Conclusion}
In the present work, we have investigated the dynamics of discord and entanglement of two atoms in Lorentzian reservoirs at zero and nonzero temperature by using the time-convolutionless master-equation method. We find that, the discord and the entanglement markedly relies on the non-Markovian effect, the detuning and the temperature as well as the atomic initial states. The entanglement is more robust than the discord when $t<t_{cri}$ while the discord is more robust than the entanglement when $t>t_{cri}$, and the value of $t_{cri}$ relies on the initial states, the temperature, the detuning and the non-Markovian effect. The discord for $|\phi(0)\rangle$ is always more robust than that for $|\psi(0)\rangle$, specially, this difference is very evident at zero temperature, and likewise for the entanglement. Nonzero temperature can accelerate the decays of discord and entanglement and induce the entanglement sudden death. More importation, if there is only the detuning or only the non-Markovian effect, both the discord and the entanglement will monotonously and rapidly decline to zero. However, if the non-Markovian effect and the detuning are present simultaneously, due to the memory and feedback of the non-Markovian reservoir, the discord and the entanglement will oscillate damply and the discord and the entanglement can be effectively protected even at nonzero temperature by increasing the detuning and the non-Markovian effect. These result will be useful in quantum computation and quantum information processing.

The current experimental technologies\cite{Raimond} show that our proposals have a certain feasibility. For example, a circular Rydberg atom with the two circular levels with principal quantum numbers 51 and 50 which are called $|e\rangle$ and $|g\rangle$ respectively, the $|e\rangle\Leftrightarrow|g\rangle$ transition is at 51.1 GHz corresponding to the atomic decay rate $\gamma_{0}=33.3$Hz. Indeed, in the above discussions, the detuning $\delta=2\gamma_{0}=66.6$Hz is very small so that it could be realized by Stark-shifting the frequency with a static electric field. The typical Stark shift is about 200kHz \cite{Hagley}, which is far more than 66.6Hz. This shift is therefore large enough to effectively preserve the discord and the entanglement. Moreover, in cavity QED experiments, ultrahigh finesse Fabry-Perot super-conducting resonant cavities with quality factors $Q=4.2\times10^{10}$, corresponding to the spectral width $\lambda=7$Hz, have been realized\cite{Kuhr}. These values correspond to $\lambda/\gamma_{0}\approx0.2$ which represents a good non-Markovian regime. Moreover, from Fig.5 and Fig.6, we may know that, the discord and the entanglement can be effectively protected when $\gamma_{0}t\leq40$(meaning a corresponding time on the scale of seconds) and $\frac{k_{B}T}{\hbar\omega_{0}}=1$(meaning the order of $T\sim10^{-8}$K), these conditions may be realized in the current experimental technologies.

\centerline{\hbox to 8cm{\hrulefill}}

\end{CJK*} 
\end{document}